\documentstyle[prl,aps]{revtex}
\begin{document}

\draft
\twocolumn[\hsize\textwidth\columnwidth\hsize\csname @twocolumnfalse\endcsname

\title{Electronic structure of Ca$_{1-x}$Sr$_x$VO$_3$: a tale of
two energy-scales}

\author{K. Maiti,$^{\star,}$\cite{tifr} D.D. Sarma,$^{\star,}$\cite{jnc} M.J. Rozenberg,$^{\S}$
I.H. Inoue,$^{\|,}$\cite{isaoad} H. Makino,$^{\P}$ O.
Goto,$^{\sharp}$ M. Pedio,$^{\dag}$ and R. Cimino$^{\ddag}$}

\address{$^{\star}$Solid State and Structural Chemistry Unit, Indian Institute of
Science, Bangalore 560012, INDIA \\ $^{\S}$Departamento de
F\'{\i}sica, Universidad de Buenos Aires, Ciudad Universitaria,
Pab. I, (1428) Buenos Aires, ARGENTINA \\ $^{\|}$PRESTO-JST and
Electrotechnical laboratory, Tsukuba 305-8568, JAPAN\\
$^{\P}$Institute for Materials Research, Tohoku University, Sendai
980-8577, JAPAN\\ $^{\sharp}$NEC Corporation, Nakahara-ku, Kawasaki
211-8666, JAPAN\\ $^{\dag}$Istituto Struttura della Materia -
C.N.R., V. Fosso del Cavaliere 100, I-00133 Roma, ITALY\\
$^{\ddag}$INFN, Sezione di Trieste, laboratori Area di Ricerca,
Padriciano 99, 34012 Trieste, ITALY}


\maketitle

\begin{abstract}
We investigate the electronic structure of Ca$_{1-x}$Sr$_x$VO$_3$
using photoemission spectroscopy. Core level spectra  establish an
electronic phase separation at the surface, leading to distinctly
different surface electronic structure compared to the bulk.
Analysis of the photoemission spectra of this system allowed us to
separate the surface and bulk contributions. These results help us
to understand properties related to two vastly differing
energy-scales, namely the low energy-scale of thermal excitations
($\sim\!k_{\rm B}T$) and the high-energy scale related to Coulomb
and other electronic interactions.
\end{abstract}


]

The electronic structure of strongly correlated transition metal
oxides has attracted a great deal of attention both
theoretically\cite{review} and experimentally\cite{RMP} due to
many exotic properties exhibited by these systems such as high
temperature superconductivity and colossal magnetoresistance. In
order to investigate such issues, photoemission spectroscopy has
been extensively employed due to its ability to probe the
electronic structure directly. While this technique is highly
surface sensitive as observed in rare earth
intermetallics\cite{RE}, its extensive use to understand the bulk
properties of transition metal (TM) oxides\cite{ddprl} is based on
the implicit assumption of very similar electronic structures at
the surface and in the bulk. We observe a spectacular failure of
this assumption in Ca$_{1-x}$Sr$_{x}$VO$_{3}$.

Ca$_{1-x}$Sr$_{x}$VO$_{3}$ is a solid solution of CaVO$_3$ and
SrVO$_3$ where the bandwidth $W$ can be systematically controlled
due to a buckling of the V-O-V bond angle from $\sim\!180^\circ$
in SrVO$_3$ to $\sim\!160^\circ$ in CaVO$_3$ \cite{inoueprb}.
Thus, Ca$_{1-x}$Sr$_x$VO$_3$ is ideally suited for the systematic
study of the competition between local interactions and itineracy,
which leads to several strong correlation effects. This system is
arguably the simplest strongly correlated transition metal oxide,
since it remains paramagnetic down to the lowest temperature
measured so far ($T$\,=\,50\,mK), has typical Fermi liquid
behavior and has nominally just one conduction electron per site
of V$^{4+}$. Despite these facts, important aspects of its
fundamental physics remain unclear, particularly in terms of its
contrasting high-energy spectroscopic and low-energy thermodynamic
properties\cite{review,marcelo}. The spectroscopic properties and
the thermodynamic properties belong to vastly different energy
scales: the former corresponds to a high energy (typically
10~$\sim$~10$^3$~eV) perturbation to the system, while the latter
probes electrons typically within $k_BT$ ($\sim$~1~meV) of $E_F$.
There is indeed {\it a-priori} no reason to believe that the same
model physics will be valid in both the regimes.

In this study, we observe a strong dependence of the photoemission
spectra from Ca$_{1-x}$Sr$_x$VO$_3$  with the escape depth
$\lambda$ of the photoelectrons, siginifying very different
surface and bulk electronic structures. The core level spectra
exhibit an electronic phase separation at the surface, possibly
due to an enhanced correlation effect and leading to a distinctly
different surface electronic structure compared to that of the
bulk. We present a method to separate the surface and bulk
contributions from the total spectrum for any given composition.
From the analysis of the bulk spectra, we deduce the values of the
electronic specific heat coefficient $\gamma$, which agree well
with the experimentally observed values. Thus, this study provides
the first coherent understanding encompassing the experimental
spectra and the measured thermodynamical properties, employing
current theoretical approaches for strongly correlated systems.

Single crystalline samples of Ca$_{1-x}$Sr$_x$VO$_3$ were prepared
by floating zone method and characterized by $x$-ray diffraction,
Laue photography and thermogravimetric analysis as described
elsewhere \cite{inoueprb}. The characterizations exhibit the
samples to be stoichiometric (error bar $<$~1\%), homogeneous and
single phasic. The $x$-ray photoemission (XP) measurements were
carried out with a monochromatized Al $K\alpha$ source with a
total resolution of 0.45~eV and the resolution for the ultraviolet
photoemission (UP) measurements were 80~meV. Sample surfaces were
cleaned by periodical scrapings and the cleanliness was confirmed
by the absence of C 1$s$ feature and the oxygen 1$s$ impurity
feature. Experiments were carried out at 120~K at a pressure of
2$\times$10$^{-10}$~mbar. A large acceptance angle
($\pm$~10$^\circ$) along with the scraped surface allow us to
probe the angle integrated spectral functions which was further
confirmed by the reproducibility of the spectra after each trial
of scraping.

In Fig.~1, we show the V 2$p_{3/2}$ core level spectra from
Ca$_{1-x}$Sr$_x$VO$_3$ for various values of $x$. It is evident
from the figure that even for CaVO$_3$ and SrVO$_3$, it does not
have the single peak structure, expected from a homogeneous single
phase V$^{4+}$ compound; instead, three clear features can be seen
(as marked) for every composition. Least-squared error analysis
showed that the positions of the components, as well as the
corresponding full widths at half maximum (FWHM) remain the same
across the series. Most significantly, the intensity ratio between
the first and the last components are always approximately in the
ratio of $1 : 1$. We find that the energy separations and the
spectral widths of these three components agree with the V
2$p_{3/2}$ signals from V$_2$O$_3$, VO$_2$ and V$_2$O$_5$,
suggesting that the peaks 1, 2 and 3 arise from V$^{3+}$, V$^{4+}$
and V$^{5+}$ entities. The equal intensities of the V$^{3+}$ and
V$^{5+}$ signals in these stoichiometric compounds suggest that a
fraction of the V$^{4+}$ ions spontaneously phase separate, $2{\rm
V}^{4+} \rightarrow {\rm V}^{3+} + {\rm V}^{5+}$, maintaining the
charge balance. By changing the photon energy between Al $K\alpha$
($1486.6$\,eV) and Mg $K\alpha$ ($1253.6$\,eV), thereby modestly
changing the surface sensitivity of the technique, we find that
with increasing surface sensitivity, the relative intensity of
V$^{4+}$ signal reduces in all the cases, as shown in the inset of
Fig.1 in the case of CaVO$_3$. Interestingly, the three components
simulating the Al~$K\alpha$ spectrum, reproduce also the
Mg~$K\alpha$ spectrum remarkably well with a smaller intensity of
V$^{4+}$ component \cite{f1}. This shows that the electronic phase
separation of V$^{4+}$ to V$^{3+}$ and V$^{5+}$ occurs at the
sample surface. Such a thing can occur {\it via} two distinctly
different mechanisms, namely due to the presence of a negative $U$
\cite{anderson,varma,sarma}, or due to the presence of strong
correlation effects \cite{Dagotto}. We believe the latter to be
the driving force in the present case, as discussed later.

It is thus clear that the different charge states of V at the
surface and in the bulk will lead to significantly different
electronic structures corresponding to the surface and the bulk.
This is evident in Fig. 2 where we show the valence band spectra
of CaVO$_3$ and SrVO$_3$ at He {\scriptsize I} (21.2~eV), He
{\scriptsize II} ($40.8$\,eV) and Al $K\alpha$ photo-excitations.
All the spectra are shown after subtracting the tail of the O 2$p$
band appearing at higher binding energies \cite{foot0}. We also
show the XP spectral function obtained from LSDA band structure
calculations for CaVO$_3$ by solid line in the figure. The
calculated spectrum exhibit {\it only one} feature for the V 3$d$
emissions at the Fermi energy, $E_{\rm F}$ representing the
delocalized conduction electrons and usually termed as coherent
feature. All the experimental spectra exhibit a second feature
centered at about 1.5 eV in addition to the coherent feature. This
feature is normally termed as incoherent feature being the
spectral signature of the lower Hubbard band (LHB) and corresponds
to electron states essentially localized due to electron
correlations. The relatively surface-sensitive He {\scriptsize I}
and {\scriptsize II} spectra in both cases are in agreement with
previously published results\cite{inoueprl}, showing a weaker
coherent feature with a main feature centered at $\sim$1.5~eV.
However, the more bulk-sensitive Al~$K\alpha$ excited spectra
exhibit much larger coherent features compared to the UP spectra.
While there is a large difference in the change in photoemission
cross sections for the photo-excitations from O 2$p$ and V 3$d$
states at these photon energies, this matrix element effects have
negligible influence in the small energy window studied here
contributed solely by the V 3$d$ states. This has been explicitly
verified in a similar system, LaVO$_3$\cite{kmprb}. Thus, these
spectral modifications establish that the electronic structures
near the surface and in the bulk are significantly different in
these compounds.

One can go beyond this qualitative discussion and present a method
to separate the intrinsic surface and bulk contributions out of
the total photoemission spectrum. We note that the total spectrum
$\rho(\omega)$ at any given photon energy can be expressed as;
$  \rho(\omega) = (1 - e^{-d/\lambda})\rho^s(\omega) +
  e^{-d/\lambda}\rho^b(\omega)$,
where $\rho^s(\omega)$ and $\rho^b(\omega)$ denote the surface and
bulk responses, $d$ is the thickness of the surface layer and
$\lambda$ is the mean free path of photoelectrons. Once the value
of $d/\lambda$ is known for two spectra (for example XPS and
He~{\scriptsize I}), one can obtain $\rho^{s}(\omega)$ and
$\rho^{b}(\omega)$ analytically.

The intensity ratios from the surface (components 1 and 3) and the
bulk (component 2) in the V 2$p_{3/2}$ spectra, which are
determined at each $x$ from the spectral decomposition, yield a
value of $d/\lambda = 0.65$ for V 2$p$ electrons with a kinetic
energy $E$ of about 965\,eV ($\sim$ $h\nu$ - BE). The simulation
of the Mg\,$K\alpha$ spectrum ($E~\sim$~735\,eV) of CaVO$_3$
represented by the solid line in the inset of Fig.1 in terms of
the three components results in $d/\lambda = 0.76$. Considering
$\lambda \propto \sqrt{E}$ in the high energy limit\cite{penedpt},
$\lambda_{Mg}/\lambda_{Al}$ is expected to be $\sqrt{735/965} =
1.15$, whereas the spectral analysis suggest
$\lambda_{Mg}/\lambda_{Al}$=0.76/0.65=1.17, in very good agreement
with the expected value, providing confidence in our analysis.
$d/\lambda$ for the valence electrons ($E\cong\!1480$\,eV) is
estimated to be $0.52\pm0.05$ for the entire series, a narrow
variation of less than 10\,\% across the series indicating a
reliable estimate of this parameter. We also need to estimate the
quantity, $d/\lambda_{\hbox{\tiny He I}}$ for the valence band
spectrum excited with He {\scriptsize I} radiation. Unfortunately,
there is no universally accepted dependence of $\lambda$ on $E$ in
the low energy limit. Thus, we assume that $\lambda_{\hbox{\tiny
XPS}}/\lambda_{\hbox{\tiny He I}}$ in Ca$_{1-x}$Sr$_x$VO$_3$ is
the same as in the closely related series
Ca$_{1-x}$La$_x$VO$_3$\cite{lacavo}; and note that the final
results for $\rho^b(\omega)$ and $\rho^s(\omega)$ are not very
sensitive to this particular choice of $\lambda_{\hbox{\tiny
XPS}}/\lambda_{\hbox{\tiny He I}}=3.4$\cite{NN1}. Thus obtained
$\rho^s(\omega)$ and $\rho^b(\omega)$ are shown in the main frame
of Fig. 3. In order to ascertain the reliability of the above
procedure, we have recorded the valence band spectra of this
series for a number of different photon energies using synchrotron
radiation from the VUV beamline (Elettra, Trieste). These spectra
were successfully synthesized, as shown in the insets by solid
lines, using linear combinations of $\rho^s(\omega)$ and
$\rho^b(\omega)$ according to the equation above, thus providing
again a non-trivial check on our procedures \cite{foot1}.

$\rho^s(\omega)$ in Fig.3 for CaVO$_3$ and SrVO$_3$ are invariably
dominated by the incoherent feature, while $\rho^b(\omega)$
contains a large coherent feature with smaller, but substantial
contributions from the incoherent feature. This suggests a highly
metallic character of the bulk electronic states, while the
surface states are essentially localized. Notably, this
observation is significantly different from rare earths where only
a quantitative change was observed in terms of a modest narrowing
of the bandwidth or changing the extent of mixed valency
\cite{RE}. We note that the $\rho^b(\omega)$ is inconsistent with
the LDA DOS due to the presence of the correlation driven
incoherent feature; therefore, we have calculated the spectral
functions of the Hubbard Hamiltonian within the dynamical
mean-field theory (DMFT) that becomes exact in the limit of large
dimensions (or large lattice connectivity)\cite{review}. The DMFT
equations are solved using Iterated Perturbation Theory (IPT) on a
Bethe lattice which captures some realistic features
\cite{review}. Thus, the theoretical results depend only on two
parameters $U$ and $W$. These parameters were varied to obtain
calculated spectra in agreement with the experimentally obtained
ones for both the surface and the bulk spectral functions in each
case. The resulting theoretical results are multiplied by the
Fermi-Dirac function ($T$=120\,K) and then convoluted with the
total experimental resolution (a Gaussian with FWHM=0.45\,eV)\@.

The calculated $\rho^b(\omega)$ (solid lines) are superimposed on
the experimental data in Fig.~3, providing a remarkable agreement
in both cases. The values of $W$ simulating the spectra are 2.4~eV
and 3.2~eV in CaVO$_3$ and SrVO$_3$, respectively with $U$ = 2~eV,
similar to the results in related strongly correlated compounds
such as V$_{2}$O$_{3}$ \cite{rozenberg}. Most significantly, the
{\em same} parameter values yield for the specific heat
coefficient, which is a much lower energy probe than PES, the
values $\gamma = 3.7$ and $5.5$\,mJ\,K$^{-2}$\,mol$^{-1}$ for
SrVO$_3$ and CaVO$_3$. These are in good agreement with the
corresponding experimental values of $6.4$ and
$7.3$\,mJ\,K$^{-2}$\,mol$^{-1}$ \cite{inouejp}. We therefore
obtain for the first time a unified understanding of the physics
at two vastly different energy-scales in this strongly correlated
system, solely based on the assumption of the Hubbard model as an
effective model\cite{foot3}. It is to be noted here that $N(E_{\rm
F})$ from DMFT \cite{foot4} is about half of that obtained from
the {\it ab initio} LDA calculations. Since the self-energy within
DMFT is momentum independent due to the local nature of the
correlations, $N(E_{\rm F})$ remains unrenormalized by $U$
\cite{foot5}. This suggests that one should rule out a naive
combination of LDA and DMFT methods as a candidate for an {\it ab
initio} technique in correlated systems, which is a subject of
strong current interest.

We now briefly comment on $\rho^s(\omega)$, representing the
average surface electronic structure, arising from the V$^{3+}$
and V$^{5+}$ dominated regions. V$^{5+}$ ions have 3$d^0$
electronic configuration and therefore do not contribute any
photoemission signal over the probed energy range; thus, the
$\rho^s(\omega)$ in Fig.~3 arises entirely from surface regions
with V$^{3+}$ species. Interestingly, $\rho^s(\omega)$ of CaVO$_3$
has virtually no intensity at $E_{\rm F}$ suggesting an insulating
state, while that of SrVO$_3$ has a finite intensity at $E_{\rm
F}$ signifying a metal. Such a change can possibly be attributed
to a more distorted crystal structure of CaVO$_3$ compared to that
in SrVO$_3$ \cite{inoueprb,inoueprl}. We have simulated
$\rho^s(\omega)$ in each case within the same DMFT formalism, and
shown by the dashed lines overlapping the experimental data in
Fig. 3. The $U/W$ required to simulate these $\rho^s(\omega)$ are
1.5 and 2 for SrVO$_3$ and CaVO$_3$, respectively. Thus, there is
evidently a strong enhancement of $U/W$ at the surface compared to
$U/W$=0.63-0.83 for the bulk in the series. This marked
enhancement may be due to the reduced atomic coordination at the
surface and/or surface reconstruction, which would give rise to a
decrease in $W$ and an increase in $U$ compared to the bulk. We
believe that this enhanced correlation effect at the surface is
also the driving force for the observed electronic phase
separation \cite{Dagotto}, in a way reminiscent of the manganites;
however, one important distinction between the two is that the
latter is a doped metallic system, while Ca$_{1-x}$Sr$_x$VO$_3$ is
a nominally undoped metallic system.

In conclusion, our present work has established that the surface
electronic structure of  Ca$_{1-x}$Sr$_x$VO$_3$ is fundamentally
different from that in the bulk. The bulk electronic structure
obtained in the present study allows for the first time a unified
understanding of the low- and the high-energy scale physics of
this system within DMFT calculations based on the Hubbard model.
This technique may also give a clue to understand one basic and
important open question which still lies ahead, but possibly
within our reach: {\it i.e.}, to find out whether the standing
conflict between the theoretical and experimental results on the
systematic evolution of {\it doped} Mott-Hubbard systems can also
be resolved within the existing paradigms of electronic structure
theories.

\acknowledgments
 We thank Drs. B. Ressel, C. Comiccioli and M. Peloi for valuable
help with the synchrotron measurements. KM and DDS acknowledge
support of the DST (India). MJR acknowledges support of
Fundaci\'on Antorchas, CONICET (PID $N^o4547/96$), and ANPCYT
(PMT-PICT1855).

{\bf Figure Captions:}

Fig.1 V $2p_{3/2}$ core level spectra (open circles) of
Ca$_{1-x}$Sr$_x$VO$_3$ for various values of $x$, exhibiting the
existence of three distinct spectral features marked 1, 2 and 3.
Dashed lines and solid lines represent the three components and
the total calculated spectra, respectively. Inset shows the
comparison of V $2p_{3/2}$ spectra of CaVO$_3$ at Mg\,$K\alpha$
(solid circles) and Al\,$K\alpha$ (open circles) photon energies.
The solid line is the simulated Mg\,$K\alpha$ spectrum in terms of
three components.

Fig. 2 Experimental valence band photoemission spectra in the V
$d$ band region of CaVO$_3$ and SrVO$_3$ using three different
photon energies, 21.2 eV (solid circles), 40.8 eV (open circles),
and 1486.6 eV (+ centered circles). All the spectra are broadened
upto the resolution of XP (1486.6~eV) spectra. The solid line
represents the XP spectral function obtained from LSDA band
structure calculations.

Fig. 3 Extracted $\rho^b(\omega)$ (open circles) and
$\rho^s(\omega)$ (solid circles) obtained from the spectra shown
in Fig. 2. The solid and dashed lines superimposed on the spectra
are the results of DMFT calculations. Insets I and II show the
comparison of experimental valence band photoemission results
(open circles) of CaVO$_3$ and SrVO$_3$, respectively for
different photon energies along with the synthesized spectra
(solid lines) obtained from $\rho^b(\omega)$ and $\rho^s(\omega)$.


\begin{thebibliography}{99}
%
\bibitem[\infty]{jnc} Also in Jawaharlal Nehru Centre for
Advanced Scientific Research, Bangalore, India.\\
 Electronic address: sarma@sscu.iisc.ernet.in
%
\bibitem[a]{tifr} Present address: Department of Condensed Matter Physics
and Materials Science, Tata Institute of Fundamental Research,
Colaba, Mumbai-400 005, India.
%
\bibitem[b]{isaoad} Present Address: Correlated Electron Research Center
(CERC), AIST Tsukuba Central 4, Tsukuba 305-8562 Japan.
%
\bibitem{review}
A. Georges {\it et al.}, Rev.\ Mod.\ Phys. {\bf 68}, 13 (1996).
%
\bibitem{RMP}
M. Imada, A. Fujimori, and Y. Tokura, Rev.\ Mod.\ Phys.
{\bf 70}, 1039 (1998).
%
\bibitem{RE}
C. Laubschat {\it et al.}, Phys.\ Rev.\ Lett. {\bf 65},
1639 (1990); L.Z. Liu {\it et al.}, Phys.\ Rev.\ B {\bf 45},
8934 (1992); J.W. Allen and L.Z. Liu, Phys.\ Rev.\ B {\bf 46},
5047 (1992).
%
\bibitem{ddprl}
D.D. Sarma {\it et al.}, Phys.\ Rev.\ Lett. {\bf 75},
1126 (1995); D.D. Sarma, N. Shanthi, and Priya Mahadevan,
Phys.\ Rev.\ B {\bf 54}, 1622 (1996).
%
\bibitem{inoueprb}
I.H. Inoue {\it et al.}, Phys.\ Rev.\ B
{\bf 58}, 4372 (1998); H. Makino {\it et al.}, Phys.\ Rev.\ B {\bf
58}, 4384 (1998).
%
\bibitem{marcelo}
M.J. Rozenberg {\it et al.}, Phys.\ Rev.\ Lett.
{\bf 76}, 4781 (1996).
%
\bibitem{f1}
The reduction in the V$^{4+}$ intensity in Mg $K\alpha$
spectra compared to that in Al $K\alpha$ spectra is in quantitative agreement
with the change in surface sensitivity between the two techniques, as discussed later in
the text.
%
\bibitem{anderson}
P.W. Anderson, Phys. Rev. Lett. {\bf 34}, 953
(1975).
%
\bibitem{varma}
C.M. Varma, Phys. Rev. Lett. {\bf 61}, 2713
(1988).
%
\bibitem{sarma}
D.D. Sarma {\it et al.}, Phys. Rev. Lett. {\bf 85}, 2549 (2000).
%
\bibitem{Dagotto}
A. Moreo {\it et al.}, Phys. Rev. Lett. {\bf 84},  5568 (2000);
M.J. Rozenberg and G. Kotliar (Unpublished).
%
\bibitem{foot0}
For clarity, we only show results for the end compositions.
%
\bibitem{inoueprl}
I.H. Inoue {\it et al.}, Phys.\ Rev.\ Lett. {\bf 74},
2539 (1995); K. Morikawa {\it et al.}, Phys.\ Rev.\ B {\bf 52},
13711 (1995).
%
\bibitem{kmprb}
K. Maiti and D.D. Sarma, Phys. Rev. B {\bf 61},
2525 (2000).
%
\bibitem{penedpt}
M.P. Seah and W.A. Dench, Surf. Interface Anal.
{\bf 1}, 2 (1979).
%
\bibitem{lacavo}
K. Maiti, Priya Mahadevan, and D.D. Sarma, Phys.\
Rev.\ Lett. {\bf 80}, 2885 (1998).
%
\bibitem{NN1}
We checked the validity of our results in a wide range of
$\lambda_{\hbox{\tiny XPS}}/\lambda_{\hbox{\tiny He I}}$
between $3.0$ and $4.5$\@. Outside this range, the extracted
$\rho^b(\omega)$ and $\rho^s(\omega)$ develop unphysical
negative intensities.
%
\bibitem{foot1}
We find that the required $d/\lambda$ values interpolate
between those for the Al~$K\alpha$ and He~{\scriptsize I}
radiations, as expected.
%
\bibitem{rozenberg}
M.J. Rozenberg {\it et al.}, Phys.\ Rev.\ Lett.
{\bf 75}, 105 (1995).
%
\bibitem{inouejp}
I.H. Inoue {\it et al.}, J. Phys.\ Condens.\ Matter (London)
{\bf 10}, 11541 (1998).
%
\bibitem{foot3}
The new data for $\rho^b(\omega)$ {\em ridden of surface
contributions} is the crucial difference with respect to similar
analysis in the previous work\cite{marcelo}.
%
\bibitem{foot4}
In DMFT we assume a semi-elliptical DOS of bandwidth $W$.
%
\bibitem{foot5}
In an earlier study \protect\cite{inoueprl}, a nonlocal
selfenergy was suggested phenomenologically to reduce the value
of $N(E_{\rm F})$. However, due to the large surface
contribution to the PES spectra an unphysically large effect of
the nonlocal selfenergy was obtained.
%
\end{thebibliography}
\end{document}